\documentclass[a4paper,twoside,notoc,11pt]{JHEP3}
\pdfoutput=1
\usepackage{epsfig,multicol,slashed}
\usepackage{delarray,amsmath,bbm}

\newcommand\fverb{\setbox\pippobox=\hbox\bgroup\verb}
\newcommand\fverbdo{\egroup\medskip\noindent%
                              \fbox{\unhbox\pippobox}\ }
\newcommand\fverbit{\egroup\item[\fbox{\unhbox\pippobox}]}
\newbox\pippobox
\newcommand{\nn}{\nonumber}
\newcommand{\beq} {\begin{equation}}
\newcommand{\eeq} {\end{equation}}
\newcommand{\beqa} {\begin{eqnarray}}
\newcommand{\eeqa} {\end{eqnarray}}

\newcommand{\eg}{{\it e.g.}}

\newcommand{\lqcd}{\Lambda_{QCD}}

\newcommand{\order}[1]{${\cal O}\left(#1 \right)$}
\newcommand{\morder}[1]{{\cal O}\left(#1 \right)}
\newcommand{\eq}[1]{(\ref{#1})}
\newcommand{\fig}[1]{Fig.~\ref{#1}}

\newcommand{\inv}[1]{\frac{1}{#1}}
\newcommand{\ket}[1]{\vert{#1}\rangle}
\newcommand{\bra}[1]{\langle{#1}\vert}
\newcommand{\ave}[1]{\langle{#1}\rangle}

\newcommand{\bs}[1]{\boldsymbol{#1}}

\newcommand{\kv}{{\bs{k}}}
\newcommand{\qv}{{\bs{q}}}
\newcommand{\bv}{{\bs{b}}}

\newcommand{\qu}{{\rm q}}
\newcommand{\qb}{{\rm\bar q}}

\newcommand{\halft}{{\textstyle \frac{1}{2}}}
\newcommand{\lsim}{\lesssim}
\newcommand{\gsim}{\gtrsim}

\title{The transverse shape of the electron}
\author{Paul Hoyer and Samu Kurki\\
              Department of Physics and Helsinki Institute of
              Physics\\
              \ POB 64, FIN-00014 University of Helsinki, Finland \\
              }
\preprint{HIP-2009-30/TH}

\abstract{We study the charge density, form factors and spin distributions of the electron 
induced by its $\ket{e\gamma}$ light-front Fock state in impact parameter space.
Only transversally compact Fock states contribute to the leading behavior of the Dirac and Pauli form factors as the momentum transfer tends to infinity. Power suppressed contributions are not compact, and distributions weighted by the transverse size have end-point contributions. The Fock state conserves the spin of the parent electron locally, but the separate contributions of the electron, photon and orbital angular momentum depend on longitudinal momentum and impact parameter. The sign of the anomalous magnetic moment of the electron may be understood intuitively from the density distribution, addressing a challenge by Feynman.}

\begin{document}

\section{Introduction}

The density of quarks in hadrons as a function of their transverse position (impact parameter) $\bv$ is given by exclusive electromagnetic form factors \cite{Soper:1976jc}. In recent years this fully relativistic connection has attracted considerable interest \cite{Burkardt:2000za,Ralston:2001xs,Diehl:2002he,Miller:2007uy,Carlson:2008zc}. The densities are defined at equal Light-Front\footnote{We use the notation $x^\pm=x^0 \pm x^3$ and denote four-vectors as $x=(x^+,x^-,\bs{x})$.} (LF) time $x^+$ and integrated over the longitudinal momentum fraction $x$ of the quark. Transverse density profiles of the nucleon and deuteron have thus been determined from form factor data \cite{Miller:2007uy,Carlson:2008zc}. Generalized parton distributions (GPD's) allow to simultaneously determine the charge densities in $x$ and $\bv$. While it is difficult to extract GPD's from data on hard exclusive processes, the proton density in $(x,\bv)$-space has been studied using GPD models \cite{Burkardt:2000za}.

Here we apply the above methods to the QED electron. The structure of the electron is interesting in its own right, but it also serves as a field theory model for hadrons. Several studies have been made of the $\ket{e\gamma}$ Fock state of the electron with this in mind \cite{Brodsky:2000ii,Chakrabarti:2004ci,Brodsky:2006ku,Burkardt:2008ua}. Here we give closed forms for the electron wave functions in impact parameter, valid in $A^+=0$ gauge. The electron density has a finite width in $\bv$ even as its $x \to 1$, and the impact parameter distribution of the photon becomes infinitely broad in this limit. We determine the transverse size distribution of the Fock states that contribute to the Dirac and Pauli form factor at any $Q^2$. Only compact states contribute at high $Q^2$, but the convergence is rather slow, especially for large momentum fractions $x$ of the electron. This indicates that color transparency is more difficult to observe in exclusive form factors than in deeply exclusive meson production. In $A^+=0$ gauge the electron spin is preserved locally in $x$ and $\bv$, in the sense of expectation values: $J^z = \ave{S_{e}^z}+\ave{S_{\gamma}^z}+\ave{L_{e\gamma}^z}$. The relative magnitudes of the three expectation values have an interesting dependence on $x$ and $\bv$. 

Studies of density and spin distributions in impact parameter space will hopefully assist in gaining a more intuitive understanding of relativistic systems. The importance of this was emphasized by Feynman at the 1961 Solvay conference. Concerning the anomalous moment of the electron Feynman notes \cite{Drell:1965hg}: ``We have no physical picture by which we can easily see that the correction is roughly $\alpha/2\pi$, in fact, we do not even know why the sign is positive (other than by computing it)''. The relation between form factors and densities seems to allow an intuitive understanding of the sign of the electron's anomalous magnetic moment.

\section{Wave functions in transverse position space}

The impact parameter density $\rho_0(\bv)$ of a spin $\halft$ target with helicity $\lambda$ is defined as the Fourier transform of the electromagnetic form factor, 
\beq\label{rho0}
\rho_0(\bv) \equiv \int \frac{d^2 \qv}{(2 \pi)^2} \,  
e^{-i \, \qv \cdot \bv} \, \frac{1}{2 P^+}   
\bra{P^+, \halft\qv, \lambda}\, j^+(0) \,\ket{ 
P^+, -\halft\qv, \lambda} = \int_0^\infty \frac{d Q}{2 \pi}\, Q \, J_0(b \, Q) F_1(Q^2)
\eeq
in the Drell-Yan frame where the virtual photon momentum $q=(0^+,0^-,\qv)$. Here $F_{1}$ is the Dirac form factor, $J_{0}$ is a Bessel function and $Q^2=\qv^2$. The fact that $\rho_0(\bv)$ may be regarded as a charge density is seen by expressing it in terms of LF wave functions $\psi_{n}^\lambda(x_{i},\bv_{i},\lambda_{i})$ describing an $n$-particle Fock state at a given LF time $x^+$, with constituents carrying momentum fractions $x_{i}$, located at (relative) impact parameters $\bv_{i}$ and having helicities $\lambda_{i}$. Using the conventions of Ref. \cite{Diehl:2002he} the transverse momentum states appearing in \eq{rho0} have the Fock state expansion
\beqa
\ket{P^+,\qv,\lambda} & = & \sum_{n,\lambda_{i}}\,\prod_{i=1}^{n}\Bigl[\int_{0}^{1}\frac{dx_{i}}{\sqrt{x_{i}}}\int\frac{d^{2}\kv_{i}}{16\pi^{3}}\Bigr]16\pi^{3}\delta(1-\sum_{i} x_{i})\,\delta^{(2)}(\sum_{i} \kv_{i})\nn\\[3mm] 
& \times &\psi_{n}(x_{i},\kv_{i},\lambda_{i})\,\ket{n;\, x_{i}P^{+},x_{i}\qv+\kv_{i},\lambda_{i}}
\eeqa
and are expressed in terms of states localized at transverse position $\bv$ by a Fourier transform
\beq\label{bstate}
\ket{P^+,\qv,\lambda} = 4\pi\int d^2\bv\,e^{i\qv\cdot\bv}\,\ket{P^+,\bv,\lambda}
\eeq
Defining the Fock state expansion of the transverse position states by
\beqa\label{bexp}
\ket{P^+,\bv,\lambda} &=& \sum_{n,\lambda_{i}} \Big[\prod_{i=1}^{n}\int \frac{dx_{i}}{\sqrt{x_{i}}} \,\int 4\pi d^2\bv_{i}\Big] \delta(1-\sum_{i} x_{i})\,\frac{1}{4\pi}\delta^{(2)}(\sum_{i}x_{i}\bv_{i}) \nn\\[3mm]
&\times& \psi_{n}^{\lambda}(x_{i},\bv_{i},\lambda_{i})\,\ket{n;x_{i}P^+,\bv+\bv_{i},\lambda_{i}}
\eeqa
the wave functions in impact parameter and momentum space are related by
\beq\label{wfrel}
\psi_{n}^{\lambda}(x_{i},\bv_{i},\lambda_{i}) = \int\Bigl[\prod_{i=1}^{n}\frac{d^{2}\kv_{i}}{16\pi^{3}}\Bigr]16\pi^{3}\,\delta^{(2)}(\sum_{i}\kv_{i})\exp\Bigl(i\sum_{i}\kv_{i}\cdot\bv_{i}\Bigr)\psi_{n}^{\lambda}(x_{i},\kv_{i},\lambda_{i})
\eeq

Using the above relations the transverse density \eq{rho0} is given by a sum of absolute squares of wave functions in impact parameter space,
\beqa\label{rho0expr}
\rho_{0}(\bv) &=& \sum_{n,\lambda_{i},k}e_{k}\Bigl[\prod_{i=1}^{n}\int dx_{i}\int 4\pi d^{2}\bv_{i}\Bigr]\delta(1-\sum_{i} x_{i})\frac{1}{4\pi}\delta^{(2)}(\sum_{i} x_{i}\bv_{i}) \nn \\[3mm]
 & \times & \delta^{(2)}(\bv-\bv_{k})\, |\psi_{n}^{\lambda}(x_{i},\bv_{i},\lambda_{i})|^2
\eeqa
where $e_{k}$ is the fractional electric charge of constituent $k$ (thus quarks and antiquarks contribute with opposite signs).
For reasons of symmetry, the density $\rho_0(\bv)$ of a target with given helicity can depend only on the magnitude $b=|\bv|$ of the impact parameter. When the target is polarized in the $+x$-direction the dependence on the azimuthal angle $\phi_{b}$ is given by the Pauli form factor $F_{2}$,
\beqa\label{rhox}
\rho_{x}(\bv) & \equiv & \int\frac{d^{2}\qv}{(2\pi)^{2}}\,e^{-i\qv\cdot\bv}\frac{1}{2P^{+}}\bra{P^{+},\halft\qv,S^{x}=+\halft}\, j^{+}(0)\,\ket{P^{+},-\halft\qv,S^{x}=+\halft} \nn\\
& = & \rho_0(\bv)+\sin(\phi_{b})\int_{0}^\infty\frac{dQ}{2\pi}\frac{Q^{2}}{2m}J_{1}(bQ)F_{2}(Q^{2})
\label{rhofromff}
\eeqa
The density $\rho_{x}(\bv)$ of a transversely polarized particle may be expressed as a local charge density as in \eq{rho0expr}, with $\psi_{n}^{\lambda} \to (\psi_{n}^{\lambda=1/2}+\psi_{n}^{\lambda=-1/2})/\sqrt{2}$.

These formally exact relations are the relativistic generalizations of the traditional interpretation of the form factors in terms of a charge distribution in ordinary 3-dimensional space. The traditional analysis is valid when the constituents move non-relativistically and $Q$ is small enough that their recoil can be neglected. LF wave functions naturally appear in \eq{rho0expr} since the photon interacts with all constituents at the same LF time $x^+=t+z$.

Using the closure of Bessel functions,
\begin{equation}\label{besselclose}
\int_{0}^{\infty}d\rho\,\rho\, J_{\nu}(\alpha\rho)J_{\nu}(\beta\rho)=\frac{1}{\alpha}\,\delta(\alpha-\beta),\quad\nu>-\frac{1}{2}
\end{equation}
equations \eq{rho0} and \eq{rhox} may be inverted. The form factors are thus expressed in terms of the charge densities as
\beqa
F_{1}(Q^{2}) & = & 2\pi\int_{0}^{\infty}db\,b\, J_{0}(bQ) \rho_{0}(\bv)\nn\\
F_{2}(Q^{2}) & = & \frac{2\pi m}{Q}\int_{0}^{\infty}db\,b\, J_{1}(bQ)\left(\rho_{x}^{\phi_{b}=\pi/2}(\bv)-\rho_{x}^{\phi_{b}=3\pi/2}(\bv)\right) 
\label{f2fromrho}
\eeqa

The local $j^+(0)$ current in the matrix elements of \eq{rho0} and \eq{rhox} may be generalized\footnote{Here we consider a single flavor of quark and omit the dependence on its fractional charge $e_{k}$.} to the non-local operator of GPD's \cite{Burkardt:2000za,Ralston:2001xs,Diehl:2002he},
\beq\label{gpdop}
j^+(0) \to \int\frac{dz^-}{8\pi} e^{ixP^+z^-/2}\, \qb(0^+,-\halft z^-,\bs{0}_{\perp})\gamma^+ \qu(0^+,\halft z^-,\bs{0}_{\perp})
\eeq
This allows the transverse densities to be measured as a function of the longitudinal momentum fraction $x$ of the struck quark. Thus $\rho_{0}(\bv) \to \rho_{0}(x,\bv)$, with
\beqa\label{qxb}
\rho_{0}(x,\bv) &=& \int \frac{d^2 \qv}{(2 \pi)^2} \,  
e^{-i \, \qv \cdot \bv} \int\frac{dz^-}{8\pi} e^{ixP^+z^-/2}\nn\\
&\times& \bra{P^+, \halft\qv, \lambda}\, \qb(0^+,-\halft z^-,\bs{0}_{\perp})\gamma^+ \qu(0^+,\halft z^-,\bs{0}_{\perp}) \,\ket{P^+, -\halft\qv, \lambda}\nn\\[4mm]
&=& \sum_{n,\lambda_{i},k}\prod_{i=1}^{n}\Bigl[\int dx_{i}\int4\pi d^{2}\bv_{i}\Bigr]\delta(1-\sum_{i}x_{i})\frac{1}{4\pi}\delta^{(2)}(\sum_{i}x_{i}\bv_{i})\nn\\[3mm]
&\times&\delta^{(2)}(\bv-\bv_{k})\delta(x-x_{k})\,|\psi_{n}^{\lambda}(x_{i},\bv_{i},\lambda_{i})|^{2}
\eeqa
which differs from \eq{rho0expr} only through the constraint $\delta(x-x_{k})$ that the momentum fraction $x_{k}$ of the struck quark be equal to $x$. The same expression applies to $\rho_{x}(x,\bv)$ with $\psi_{n}^{\lambda} \to (\psi_{n}^{\lambda=1/2}+\psi_{n}^{\lambda=-1/2})/\sqrt{2}$.

We may now apply the above expressions to the $\ket{e\gamma}$ Fock state of a QED electron localized at $\bv=0$. According to \eq{bexp} $x_{e}\equiv x = 1-x_{\gamma}$, $\bv_{e} \equiv \bv = -(1-x)\bv_{\gamma}/x$ and we denote $m_{e} \equiv m$. The impact parameter wave functions are expressed in terms of the momentum space ones as in \eq{wfrel},
\beq\label{bkwf}
\psi_{\lambda_{e}\lambda_{\gamma}}^\lambda(x,\bv) = \int\frac{d^2\kv}{16\pi^3}\exp\Big[i\frac{\bv\cdot\kv}{1-x}\Big]\psi_{\lambda_{e}\lambda_{\gamma}}^\lambda(x,\kv)
\eeq
Using the $A^+=0$ gauge momentum space wave functions given in \cite{Brodsky:2000ii,Brodsky:1980zm} and the identity
\beq
\int_{0}^\infty \frac{t^{\nu+1}J_{\nu}(at)}{(t^2+z^2)^{\mu+1}}dt = \frac{a^{\mu}z^{\nu-\mu}}{2^\mu\Gamma(\mu+1)}\,K_{\nu-\mu}(az)
\eeq 
we have, denoting $\bv= b (\cos \phi_b, \sin \phi_b)$ and $\lambda=\pm \halft \equiv \pm$,
\beqa\label{wfs}
\psi_{+\frac{1}{2}+1}^{+}(x,\bv) & = & \psi_{-\frac{1}{2}-1}^{-\ \dagger}(x,\bv) = -i\,\frac{em\sqrt{1-x}}{4\sqrt{2}\,\pi^{2}}\,e^{-i\phi_{b}}K_{1}(mb)\nn\\[2mm]
\psi_{+\frac{1}{2}-1}^{+}(x,\bv) & = & \psi_{-\frac{1}{2}+1}^{-\ \dagger}(x,\bv) =\ i\frac{em\sqrt{1-x}}{4\sqrt{2}\,\pi^{2}}\,x\,e^{+i\phi_{b}}K_{1}(mb)\nn\\[2mm]
\psi_{-\frac{1}{2}+1}^{+}(x,\bv) & = & \psi_{+\frac{1}{2}-1}^{-}(x,\bv) = -\frac{em\sqrt{1-x}}{4\sqrt{2}\,\pi^{2}}\,(1-x)K_{0}(mb) \label{flipwfs}\nn \\[2mm]
\psi_{-\frac{1}{2}-1}^{+}(x,\bv) & = & \psi_{+\frac{1}{2}+1}^{-}(x,\bv) = 0
\eeqa

It is interesting to notice that there is no explicit factor of $m$ associated with wave functions in which the electron helicity flips. The $m$-dependence appears through the index of the Bessel $K$-function, which also indicates the value of the orbital angular momentum $L_{e\gamma}$ between the electron and the photon.

%
\EPSFIGURE[ht]{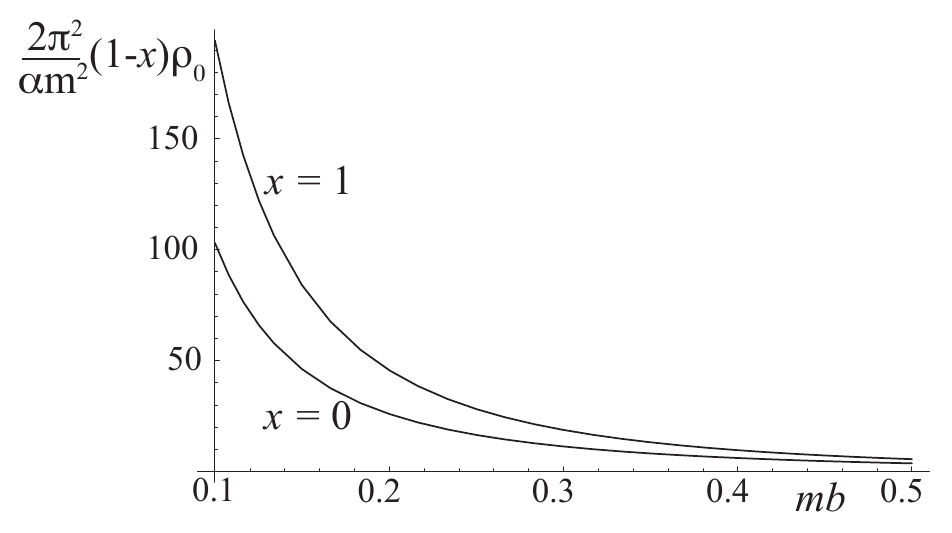,width=.8\columnwidth}{$b$-dependence of $(1-x)\rho_{0}(x,\bv)\,2\pi^2/\alpha m^2$ at $x=0$ and $x=1$. Note that the $y$-axis is at $b=0.1/m$.
\label{b-shape}}
%

The $\ket{e\gamma}$ wave functions determine the electron densities at \order{\alpha} in QED according to \eq{qxb},
\beqa\label{rhoqed}
\rho_{0}(x,\bv) &=& \frac{\alpha m^{2}}{2\pi^{2}}\Bigl[\frac{1+x^{2}}{1-x}K_{1}^{2}(mb)+(1-x)\, K_{0}^{2}(mb)\Bigl]\nn\\[3mm]
\rho_{x}(x,\bv) &=& \rho_{0}(x,\bv) + \frac{\alpha m^{2}}{\pi^{2}}x\,\sin(\phi_{b})\, K_{0}(mb)K_{1}(mb)
\eeqa
Since $K_{n}(z) \simeq e^{-z}\sqrt{\pi/2z}$ for any $n$ as $z \to \infty$ the $b$-dependence becomes independent of $x$ (and $\phi_{b}$) when $b \gg 1/m$. For $z \to 0$,  $K_{0}(z) \simeq \log(1/z)$ while $K_{1}(z) \simeq 1/z$ and so the densities become more peaked at low $b \ll 1/m$ as $x \to 1$. This illustrated in \fig{b-shape} where we plot the  $b$-dependence of $(1-x)\rho_{0}(x,\bv)\,2\pi^2/\alpha m^2$ at $x=0$ and $x=1$. We note that the electron nevertheless has a broad distribution in impact parameter for all $x$. Due to the constraint $\sum_{i}x_{i}\bv_{i}=0$ in \eq{bexp} we might have expected $\rho_{0}(x,\bv) \to \delta^2(\bv)$ for $x\to 1$ (where $x_{\gamma}=1-x \to 0$). However, the photon impact parameter $\bv_{\gamma} =-x\bv/(1-x) \to \infty$ in this limit, allowing $\rho_{0}(x,\bv)$ to remain wide. This feature is specific to QED. For hadrons confinement imposes $\bv_{\gamma} \lsim 1/\lqcd$, forcing $\bv\to 0$ as $x\to 1$.

The $\sin\phi_{b}$-dependence in $\rho_{x}(x,\bv)$ arises from one unit of angular momentum. Hence it is natural that this part of the density is less peaked (by a factor $b$) than $\rho_{0}(x,\bv)$ as $\bv\to 0$, and vanishes for $x \to 0$ (since $\bv_{\gamma} \to 0$ in this limit).

\section{Impact parameter analysis of form factors}

According to \eq{f2fromrho} the anomalous magnetic moment of the electron is given by its charge density as
\beq\label{anom}
F_{2}(0) = \pi m\int_{0}^{\infty}db\,b^{2}\left(\rho_{x}^{\phi_{b}=\pi/2}(\bv)-\rho_{x}^{\phi_{b}=3\pi/2}(\bv)\right)
\eeq
This is an exact relation, valid to all orders in $\alpha$. There is an intuitive, classical argument  \cite{Burkardt:2000za} for the density difference in \eq{anom} (and hence the anomalous magnetic moment) to be positive. The density $\rho_{x}$ given by \eq{rhox} is the matrix element of the $j^+ = j^0+j^3$ current at $x^+=0$ of a particle with spin along the positive $x$-axis. As illustrated in \fig{anomom}, for a classical spinning body $j^3(y>0)=-j^3(y<0)>0$. Consequently we may expect that $\rho_{x}^{\phi_{b}=\pi/2}(\bv)-\rho_{x}^{\phi_{b}=3\pi/2}(\bv)>0$ in \eq{anom}.
%
\EPSFIGURE[h]{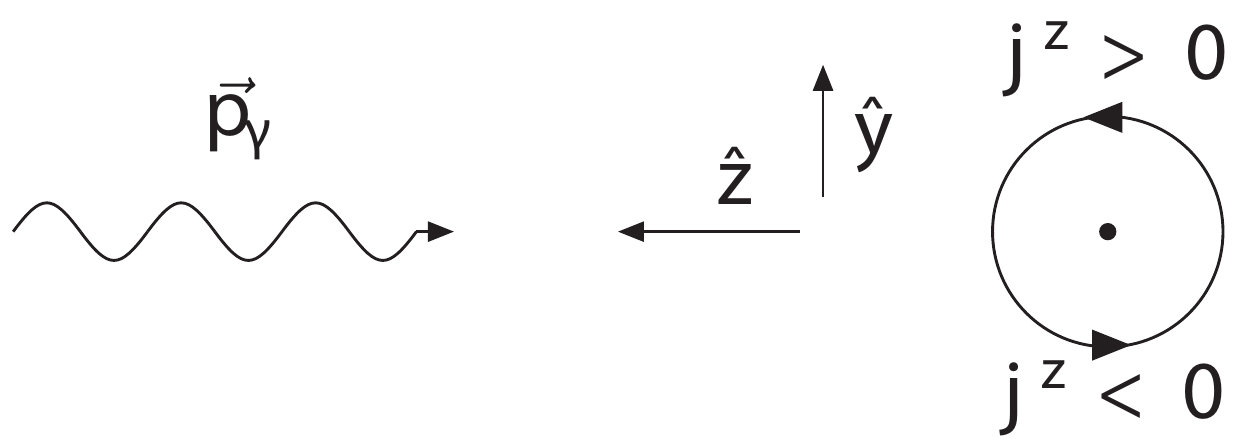,width=.7\columnwidth}{The current $j^z$ of a classical spinning particle with $L^x >0$ (out of the plane) is positive for $y>0$ and negative for $y<0$. Photons with momenta $p_{\gamma}^z < 0$ thus measure a current $j^+ = j^0+j^3$ at equal $x^+=t+z$ which is larger for $y>0$ than for $y<0$. Figure adapted from \cite{Burkardt:2000za}.
\label{anomom}}
%

Intuitive arguments like this are not reliable by themselves\footnote{See \cite{Drell:1965hg,Welton:1948zz} for earlier, more quantitative discussions of the sign and magnitude of $g-2$.}, but as shortcuts to known correct results they may help in picturing the dynamics of relativistic interactions, in the spirit of Feynman's challenge \cite{Drell:1965hg}. Analogous arguments have been used to interpret the sign of single spin (Sivers) asymmetries observed in semi-inclusive DIS \cite{Burkardt:2002ks}.

Using the QED expressions \eq{rhoqed} for the transverse densities in the general relation \eq{f2fromrho} the Pauli form factor of the electron is at \order{\alpha} given by
\beqa
F_{2}(Q^2) &=&\frac{4\alpha m^3}{\pi Q}\int_{0}^{1}dx\, x\,\int_{0}^{\infty}db\,b\, J_{1}(b\,Q)K_{0}(mb)K_{1}(mb) \label{f2b}\\[3mm]
&=& \frac{2\alpha m^2}{\pi} \inv{Q\sqrt{Q^2+4m^2}}\log\left[\inv{2m}\left(\sqrt{Q^2+4m^2}+Q\right)\right]  \label{f2q}
\eeqa
where \eq{f2q} is the standard textbook \cite{IZ} result obtained from the loop integral. 
Since the Pauli form factor is UV finite the integral in \eq{f2b} is regular and the 
equivalence of \eq{f2b} and \eq{f2q} may be verified numerically. At $Q^2=0$ \eq{f2b} gives the standard \order{\alpha} result for the anomalous moment,
\beq\label{anomqed}
F_{2}(0) = \frac{2\alpha m^{3}}{\pi} \int_{0}^{1}dx\, x\,\int_{0}^{\infty}dbb^{2} K_{0}(mb)K_{1}(mb) = \frac{\alpha}{2\pi}
\eeq

The representation \eq{f2b} of $F_{2}(Q^2)$ allows to investigate the transverse size distribution of $\ket{e\gamma}$ Fock states which contribute to the Pauli form factor of the electron at any $Q^2$. Analogous studies have been made in terms of GPD models for the nucleon \cite{Diehl:2004cx,Guidal:2004nd}. For $Q^2 \to \infty$ at fixed $x$ it is generally expected that only compact Fock states with $b \lsim 1/Q$ contribute to form factors, based on the uncertainty principle and in accordance with the Brodsky-Lepage dynamics \cite{Lepage:1980fj} of exclusive form factors. At the endpoints $x \simeq 0,1$ on the other hand contributions of large transverse size may become important. The transverse size of the $\ket{e\gamma}$ Fock state is given by $|\bv_{e}-\bv_{\gamma}|=b/(1-x)$, and thus increases as $x \to 1$.

If the leading contribution to $F_{2}(Q^2)$ in the limit $Q^2 \to \infty$ comes from small impact parameters ($b\to 0$) the Bessel $K$-functions in \eq{f2b} are evaluated at small argument, thus $K_{0}(mb) \simeq \log(1/mb)$ and $K_{1}(mb) \simeq 1/mb$. Then, with $t=bQ$,
\beqa \label{lowb}
F_{2}(Q^2,b\to 0) &\simeq& \frac{2\alpha m^2}{\pi Q^2}\int_{0}^\infty dt J_{1}(t) \log\left(\frac{Q}{m}\inv{t}\right) = \frac{\alpha m^2}{\pi Q^2} \log\left(\frac{Q^2}{m^2}\right)\left[1+\morder{\frac{1}{\log(Q^2/m^2)}}\right]\nn\\[2mm]
\eeqa
which agrees with the leading $Q^2 \to \infty$ behavior of the explicit expression \eq{f2q}. This confirms that only small impact parameters contribute to the Pauli form factor of the electron at high $Q^2$. The convergence to small $b$ is fairly slow. As shown in \fig{conv} the integral in \eq{f2b} gets contributions of \order{10\%} up to $b \lsim 0.5/m$ even when $Q^2 = 100\, m^2$.

%
\EPSFIGURE[h]{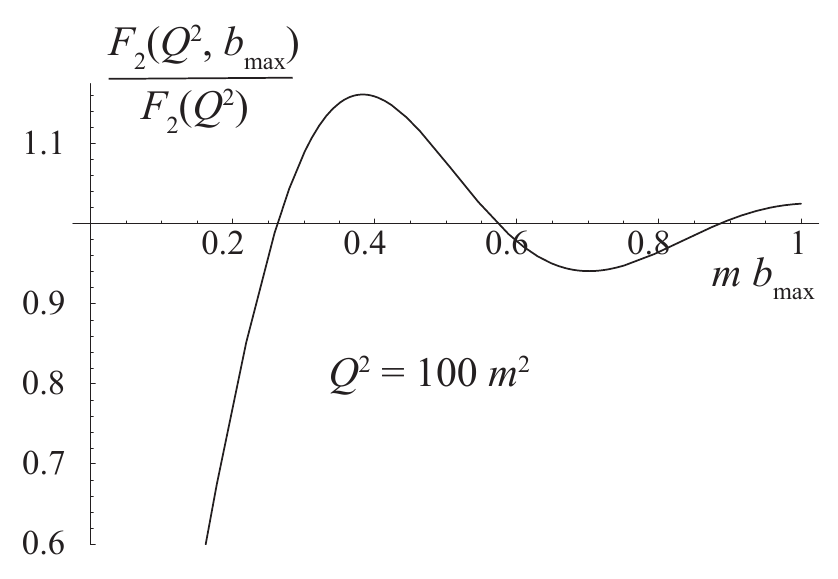,width=.7\columnwidth}{Dependence of the representation \eq{f2b} for $F_{2}(Q^2)$ to a restriction of the integral to $b \leq b_{max}$, when $Q^2=100\, m^2$.
\label{conv}}
%

The \order{\alpha} expression for the Dirac form factor $F_{1}(Q^2)$ of the electron is obtained from \eq{f2fromrho} and \eq{rhoqed}:
\beq
F_{1}(Q^2) =\frac{\alpha m^2}{\pi}\int_{0}^{1}dx\int_{0}^{\infty}db\,b\, J_{0}(b\,Q)\left[\frac{1+x^2}{1-x}K_{1}^2(mb)+(1-x)K_{0}^2(mb)\right] \label{f1}
\eeq
The usual IR singularity appears through the denominator $1-x$, and the UV behaviour as the logarithmic singularity at $b=0$, since $K_{1}(mb) \simeq 1/mb$ for $b \ll 1/m$. The term $\propto K_{0}^2$ in the brackets is regular, and arises from the wave function where the electron flips its helicity.

Renormalization of the UV singularity in \eq{f1} leaves a pointlike contribution $(b \lsim 1/Q)$ to $F_{1}(Q^2)$, which at large $Q^2$ is of \order{Q^0} (up to logarithms). The $b \lsim 1/Q$ contribution to $F_{1}(Q^2)$ from the term where the electron helicity flips is suppressed by a factor $m^2/Q^2$. However, the low $b$ approximation for this term gives an integral which diverges for large $b \equiv u/m$,
\beq\label{lowb1}
F_{1}(Q^2,b\to 0) \simeq \frac{\alpha}{\pi}\int_{0}^1 dx\int_{0}^\infty du J_{0}\left(u\frac{Q}{m}\right)\left[\frac{1+x^2}{1-x}\inv{u}+ (1-x)u\log^2u\right]
\eeq
Thus the power suppressed (higher twist) contributions to the electron's Dirac form factor cannot be restricted to small impact parameters $(b \to 0)$ at any $Q$.

As we noted above, the impact parameter of the photon, and hence also the transverse size of the $\ket{e\gamma}$ Fock state, increases for $x\to 1$. A similar behavior is expected for quark fluctuations $q \to qg$ in hadrons. This could affect the color transparency of Fock states recoiling in a nucleus. The rescattering amplitude is proportional to the dipole moment (transverse size) of the fluctuation, $r_{\perp} = b/(1-x)$. Thus suppression of rescattering requires that the Fock states which contribute to the form factor remain compact even when weighted with $r_{\perp}$. At fixed $x<1$ the $r_{\perp}$-weighted leading twist contribution to the $b$-integral in \eq{f1} is similar to the one in the Pauli form factor, which we found to be dominated by compact Fock states. However, the $x$-integral $\sim \int dx/(1-x)^2$ is linearly divergent at $x=1$, suggesting possible contributions from this end-point.

The fluctuating phase evident in \fig{conv} originates from the $J_{1}(bQ)$ Bessel function in \eq{f2b} and reflects the phase of a photon of wavelength $1/Q$ probing transverse distances of \order{b}. It is interesting to compare this case of exclusive form factors ($x_{bj}=1$) with the case of Deeply Exclusive Meson Production, $\gamma^*(Q^2)A \to M A$ at low $x_{bj}$. In the latter situation the relevant size of the Fock states of the meson $M$ is determined by the wave function of a virtual photon with large longitudinal momentum, which decreases exponentially for $b \gsim 1/Q$ according to a Bessel $K_{0,1}(bQ)$-function (see, \eg, the Appendix of \cite{Brodsky:1996nt}). Hence the Fock state size distribution stays limited even when multiplied by an arbitrary power of $b$. In fact, the positive evidence for Color Transparency indeed comes from meson electroproduction \cite{Airapetian:2002eh}, whereas a measurement of the nucleon form factor in a nuclear environment failed to see a signal for color transparency \cite{Garrow:2001di}.

\section{Distribution of the electron spin in $x$ and $\bv$}

It is interesting to study how the spin of the electron is transferred to its $\ket{e\gamma}$ Fock state. The Ji sum rule \cite{Ji:1996ek},
\beq\label{jisum}
\inv{2}\sum_{q}\Delta q +\sum_{q}L_{q}^z+J_{g}^z = \inv{2}
\eeq
expresses the $J^z=\halft$ of a proton as a sum of the total helicity $\halft\Delta q$ and orbital angular momentum $L_{q}^z$ of quarks plus the gluon spin contribution $J_{g}^z$. These contributions are related to GPD's and thus in principle measurable. Recently the $e \to e\gamma$ system was used  \cite{Burkardt:2008ua} to study the difference between \eq{jisum} and an alternative decomposition \cite{Jaffe:1989jz}.

Here we shall study how the electron spin is carried {\it locally} in $(x_{i},\bv_{i})$ by its $\ket{e\gamma}$ Fock constituents. The wave functions \eq{wfs} of $A^+=0$ gauge are appropriate for this, since the photon only has transverse polarization states. The expectation values of the electron helicity $\lambda_{e}$, the photon helicity $\lambda_{\gamma}$ and the orbital angular momentum $L_{e\gamma}^z=-i\partial_{\phi_{b}}$ are well-defined and allow a physical interpretation.

The normalization of the \order{\alpha} Fock state is proportional to the density $\rho_{0}(x,\bv)$ in \eq{rhoqed},
\beq
N \equiv \sum_{\lambda_{e},\lambda_{\gamma}}\left|\psi_{\lambda_{e}\lambda_{\gamma}}^+\right|^2 = \frac{\alpha m^{2}(1-x)}{8\pi^{3}}\Bigl[(1+x^{2})K_{1}^{2}(mb)+(1-x)^2 K_{0}^{2}(mb)\Bigl]
 =\frac{(1-x)^2}{4\pi}\rho_{0}(x,\bv)
\eeq
We define the expectation value of the electron helicity $\lambda_{e}$ in the $\ket{e\gamma}$ Fock state of a parent electron with helicity $\lambda$ as
\beq\label{expectdef}
\ave{\lambda_{e}}_{\lambda}\equiv \inv{N}\sum_{\lambda_{e},\lambda_{\gamma}}\bra{\lambda;\lambda_{e},\lambda_{\gamma}}S_{e}^z \ket{\lambda;\lambda_{e},\lambda_{\gamma}}
=\inv{2N}\sum_{\lambda_{\gamma}}\left[|\psi_{+,\lambda_{\gamma}}^{\lambda}|^2- |\psi_{-,\lambda_{\gamma}}^{\lambda}|^2\right]
\eeq
where $S_{e}^z = \halft \sigma^{z}$ is the electron spin operator.
When integrated over the electron impact parameter $\ave{\lambda_{e}}_{\lambda}$ is proportional to the spin dependent distribution $g_{1}(x)$ of the electron.
The expectation values for the photon helicity and the angular momentum are defined analogously. For an electron with positive helicity $\lambda=+\halft$ the wave functions \eq{wfs} give
\beqa
N\ave{\lambda_{e}}_{+}&=& \frac{\alpha m^{2}(1-x)}{8\pi^{3}}\inv{2}\Bigl[(1+x^{2})K_{1}^{2}(mb)-(1-x)^2 K_{0}^{2}(mb)\Bigl] \nn\\
N\ave{\lambda_{\gamma}}_{+}&=& \frac{\alpha m^{2}(1-x)}{8\pi^{3}}\Bigl[(1-x^{2})K_{1}^{2}(mb)+(1-x)^2 K_{0}^{2}(mb)\Bigl] \\
N\ave{L_{e\gamma}^z}_{+}&=& -\frac{\alpha m^{2}(1-x)}{8\pi^{3}}\Bigl[(1-x^{2})K_{1}^{2}(mb)\Bigl] \nn
\eeqa
We note that the expectation values add up the electron helicity locally in $x$ and $\bv$:
\beq 
\ave{\lambda_{e}}_{+}+\ave{\lambda_{\gamma}}_{+}+\ave{L_{e\gamma}^z}_{+}=\halft
\eeq
which could be anticipated since this relation holds separately for each wave function in \eq{wfs}. The individual spin contributions depend on $x$ and $b$, as seen in \fig{spin}.
%
\EPSFIGURE[h]{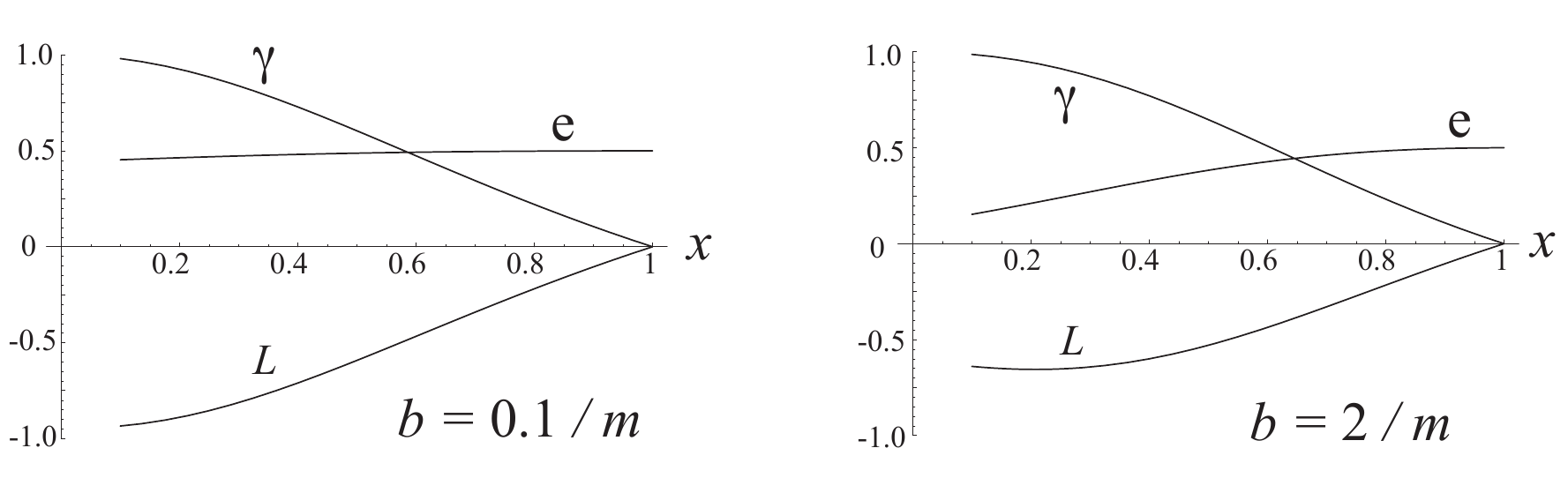,width=.9\columnwidth}{Contributions to the helicity $+\halft$ of the parent electron from the Fock state electron ($e$), the photon ($\gamma$) and the orbital angular momentum ($L$). The distributions are shown as a function of the momentum fraction $x$ of the electron, for electron impact parameters $b=0.1/m$ and $b=2/m$.
\label{spin}}
%
As $x \to 1$ the electron carries all of the spin, while the photon and orbital angular momentum $(L)$ contributions vanish. This makes the $\ket{e\gamma}$ Fock state similar to the \order{\alpha^0} bare electron, allowing virtual and real photon emissions to cancel the IR singularity. At low $x$ the electron carries the full spin for $b \to 0$, while the photon and $L$ contributions are large and cancel each other. For $x\to 0$ and $b\to\infty$ the spin is carried jointly by the photon and the orbital angular momentum.

Finally we consider the transverse spin carried by the Fock state electron,
for an electron polarized in the $x$-direction:
\beq\label{transversity}
\ave{S_{e}^x}_{x}\equiv \inv{N}\sum_{\lambda_{e},\lambda_{\gamma}}\bra{S^{x}=\halft;\lambda_{e},\lambda_{\gamma}}S_{e}^x \ket{S^{x}=\halft;\lambda_{e},\lambda_{\gamma}}
\eeq
where 
$\ket{S^{x}=\halft;\lambda_{e},\lambda_{\gamma}}=\Big[\ket{\lambda=\halft;\lambda_{e},\lambda_{\gamma}}+\ket{\lambda=-\halft;\lambda_{e},\lambda_{\gamma}}\Big]/\sqrt{2}$
and $S_{e}^x \ket{S^{x}=\halft;\lambda_{e},\lambda_{\gamma}}=\halft\ket{S^{x}=\halft;-\lambda_{e},\lambda_{\gamma}}$. When $\ave{S_{e}^x}_{x}$ is integrated over the electron impact parameter it is proportional to the transversity distribution $h_{1}(x)$ of the electron.
We find at \order{\alpha},
\beq\label{transres}
\ave{S_{e}^x}_{x}=\inv{N}\frac{\alpha m^{2}(1-x)}{8\pi^{3}}\Bigl[x K_{1}^{2}(mb)+\sin\phi_{b}(1-x) K_{0}(mb) K_{1}(mb)\Bigl]
\eeq
As seen in \fig{trans} the electron carries all of the transverse spin as $x \to 1$, while it contributes an amount $\propto x$ at low $x$. These features are qualitatively the same at all impact parameters $b$ of the electron.

%
\EPSFIGURE[h]{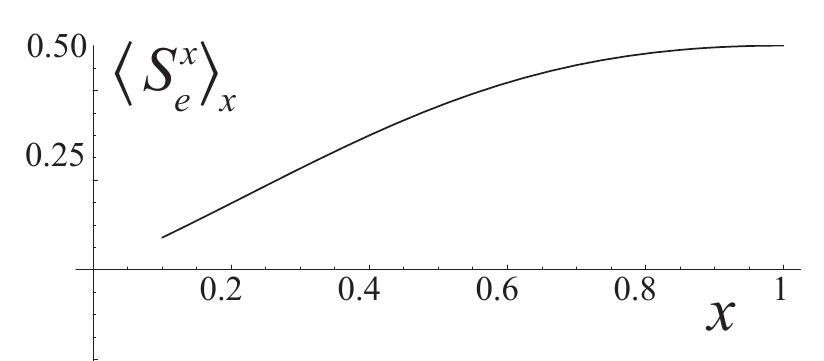,width=.7\columnwidth}{The transversity \eq{transres} of an electron as a function of its longitudinal momentum fraction $x$, integrated over the azimuthal angle $\phi_{b}$ and for $b = 1/m$.
\label{trans}}
%

\section{Discussion}

As emphasized long ago by Feynman \cite{Drell:1965hg} it is important to gain intuitive insight into the properties of relativistic systems. The understanding of the exact relation between measurable form factors and charge densities in impact parameter space \cite{Soper:1976jc,Burkardt:2000za} has been an important advance in this regard, which has allowed depicting hadron shapes \cite{Miller:2007uy,Carlson:2008zc} directly from form factor data. The extension to Generalized Parton Distributions makes it possible to study the impact parameter densities as a function of the longitudinal momentum fraction. 

Here we applied these methods to the QED electron, motivated both by an improved understanding of the electron in its own right, and regarding it as template for hadrons. We found that the impact parameter distribution of the electron in the $\ket{e\gamma}$ Fock state does not shrink to a point even as its momentum fraction $x \to 1$. This is possible because the transverse distribution of the photon grows without bound in this limit. This is qualitatively different from QCD hadrons, where confinement limits all impact distributions to $b_{i}\lsim 1/\lqcd$. Comparisons between the electron and hadron form factors cannot, however, be taken too literally since the electron carries charge, whereas hadrons are color neutral. This will affect higher order Sudakov effects induced by the emission of soft and collinear gauge bosons.

We showed that the transverse size of the Fock states which contribute to the leading order Dirac and Pauli form factors of the electron tend to zero as the momentum transfer $Q \to \infty$. However, the convergence is fairly slow and does not apply to the higher twist (power suppressed) contributions. If the distributions are weighted by the transverse size of the Fock state, as would be relevant for color transparency in a nucleus, the size distribution diverges linearly as $x\to 1$. The situation is qualitatively different in the case of meson electroproduction, where the Fock state size is determined by a photon wave function which decreases exponentially for $b \gsim 1/Q$ (except at the endpoints).

The spin of the parent electron is carried in a non-trivial way by its $\ket{e\gamma}$ constituents. Since the electron mass $m$ is the only dimensionful scale, the distributions scale in $bm$ and $x$. There is no mass dependence in the helicity flip of the electron: as the mass decreases the spin flip shifts to a correspondingly larger impact parameter. In the limit $b \to 0$ the electron constituent carries the full spin, while the photon helicity and orbital angular momentum remain large and cancel each other. The constituent electron carries all of the transverse spin of its parent as $x\to 1$, but its transversity is $\propto x$ at low values of $x$.

\acknowledgments

We are grateful for helpful discussions with Stan Brodsky. Part of this work was done at the CERN Theory Division and at ECT* (Trento), whose hospitality is highly appreciated. PH has benefitted from travel support from the Magnus Ehrnrooth foundation. SK acknowledges a PhD study grant from the Jenny and Antti Wihuri Foundation.

\end{document}